\documentclass[12pt]{JHEP3}

\let\ifpdf\relax
\usepackage{ifpdf}
\usepackage{graphics}
\usepackage{graphicx}
\usepackage{amssymb}
\usepackage{amsmath}
\usepackage{slashed}
\usepackage{cite}
\usepackage{epsfig}
\usepackage{datetime}

\newcommand{\be}{\begin{equation}}
\newcommand{\ee}{\end{equation}}
\newcommand{\bal}{\begin{align}}
\newcommand{\eal}{\end{align}}
\newcommand{\bea}{\begin{eqnarray}}
\newcommand{\eea}{\end{eqnarray}}

\def\Tr{{\rm Tr}}

\def\tr{{\rm tr}}

\def\2kn{\frac{2\pi k}{n}}

\def\sun1{{\widehat{\rm SU}(N)_1}}

\newcommand{\R}{\mathcal{R}}

\title{Topological R\'enyi and Entanglement Entropy for a 2d q-deformed $U(N)$ Yang-Mills theory and its Chern-Simons dual}

\author{Howard J. Schnitzer  \\
{\small  Martin Fisher School of Physics, Brandeis University, \\ \ \ \ \ \ Waltham, MA 02454, USA\\
}

E-mail:  \email{schnitzr@brandeis.edu}}

\preprint{
 BRX-TH-6303}

\abstract{

R\'enyi and entanglement entropies are constructed for 2d q-deformed topological Yang-Mills theories with gauge group $U(N)$, as well as the dual 3d Chern-Simons (CS) theory on Seifert manifolds. When $q=\exp[2\pi i/(N+K)]$, and $K$ is odd, the topological R\'enyi entropy and Wilson line observables of the CS theory can be expressed in terms of the modular transformation matrices of the WZW theory, $\rm{\hat{U}(N)}_{K,N(K+N)}$. If both $K$ and $N$  are odd, there is a level-rank duality of the 2d qYM theory and of the associated CS theory, as well as that of the R\'enyi and entanglement entropies, and Wilson line observables.
}

\begin{document}

\section{Introduction}

\, Entanglement entropy (EE) provides a useful tool for the study of various aspects of quantum field theory, gravitation, AdS/CFT, conformal field theory (CFT) as well as applications to condensed matter physics. The computation of EE in the presence of gauge fields poses certain difficulties, as in general it is not possible to separate the Hilbert space into the product of two Hilbert spaces without violating Gauss's law on the boundary of a bi-partition. Typical proposals to deal with this issue have been formulated in terms of boundary conditions of a lattice gauge theory\cite{Velytsky:2008rs,Buividovich:2008kq,Buividovich:2008gq,Nakagawa:2009jk,Klebanov:2011td,Eling:2013aqa,Agon:2013iva,Casini:2013rba,Donnelly:2011hn,Donnelly:2014gva,Casini:2014aia,Casini:2015dsg,Donnelly:2014fua,Donnelly:2015hxa,Huang:2014pfa,Ghosh:2015iwa,Aoki:2015bsa,Chen:2015kfa,Hung:2015fla,Radicevic:2014kqa,Radicevic:2015sza,Soni:2015yga,VanAcoleyen:2015ccp} 

However, there are special cases where the EE can be computed in the presence of continuum gauge fields. Among theses are 

\begin{enumerate}
\item[1)] The WZW $\sun1=\hat{\rm U}(N)_1/{\rm U(1)}$ theory as described by N free Dirac fermions coupled to a $\rm U(1)$ constraint gauge field\cite{1990NuPhB.332..583N,Schnitzer:2015ira,Schnitzer:2015iip},
\item[2)] Left-right entanglement entropy (LREE) for WZW models on a circle\cite{PandoZayas:2014wsa,Das:2015oha,Schnitzer:2015gpa,Brehm:2015plf},
\item[3)] 2d non-Abelian pure gauge theory\cite{Gromov:2014kia}, and 
\item[4)] 2d q-deformed Yang-Mills theory and its Chern-Simons (CS) dual \cite{Naculich:2007nc,Naculich:1990pa,Naculich:1992uf}. 
\end{enumerate}
It is the last theory which is the subject of this paper. Hopefully, these examples will give insights for a better understanding of the computation of EE in the presence of gauge fields.

The 2d q-deformed YM theory on a genus g Riemann surface is a topological theory \cite{Naculich:2007nc,Naculich:1990pa,Naculich:1992uf},\cite{Aganagic:2004js,deHaro:2005rn,Blau:2006gh,Marino:2001re,Marino:2004uf,Blau:1993hj},\cite{Kitaev:2005dm,Dong:2008ft,Wen:2016snr} which has a close connection to CS theory on a Seifert manifold. In this paper we consider the $\rm U(N)$ WZW theory defined by 
\bea
\rm{\hat{U}(N)}_{K,N(K+N)}=[\widehat{SU}(N)_K \times \hat{U}(1)_{N(K+N)}]/\mathbb{Z}_N
\eea
which is only well defined for $K$ odd. The partition function of the 2dq U(N) YM theory on the Riemann surface $\Sigma_g$ with genus $g$, is given up to an overall normalization by\cite{Naculich:2007nc,Naculich:1990pa,Naculich:1992uf}
\bea
\label{1.2}
Z_{qYM}[U(N),q,p,\theta]\sim \sum_{\R}(\dim_q \mathcal{R})^{2-2g}q^{-\frac{1}2pC_2(\mathcal{R})}e^{i\theta C_1(\mathcal{R})}
\eea
where the sum is over all representations $\mathcal{R}$ of $\rm{U(N)}$, $\dim_q \mathcal{R}$ is the q-dimension of $\mathcal{R}$, and $C_2(\R)$ and $C_1(\R)$ are quadratic and linear Casimir operators appropriate to $\R$. 

Consider the subsystem $A$ to be the union of $l$ disjoint intervals, referred to as $l$ cuts, and the $n-$ sheeted ramified cover $\Sigma_n$ of $\Sigma_g$, with Euler characteristic 
\bea
\label{1.3}
\chi(\Sigma_n)=n\chi(\Sigma_g)-2l(n-1)
\eea
where $\chi (\Sigma_g)=2-2g$ and $l\geq 1$.
Equations (\ref{1.2}) and (\ref{1.3}) provide the basic information for computing the R\'enyi entropy with the replica trick for the case we are considering.

When $q=\exp[2\pi i /(N+K)]$, and $K$ and $p$ both integers, the set of $U(N)$ representations are described by Young tableaux with no more than $K$ columns. Moreover, the quantities in (\ref{1.2}) may then be expressed in terms of the modular transformation matrices of the WZW theory. Finally, when $\theta=0$, $K$ odd and $p$ an integer, the partition function (\ref{1.2}) may be expressed as that of a $CS$ theory on the Seifert manifold $\mathcal{M}(g,p)$, where $p$ is the first Chern class of the circle bundle over $\Sigma_q$.  This last relation gives a dual description of the R\'enyi and entanglement entropies. These considerations can also be extended to certain Wilson line observables\cite{Naculich:2007nc,Naculich:1990pa,Naculich:1992uf}. 

Further, when $N$ and $K$ are both odd, the qYM theory exhibits $N \leftrightarrow K$ duality\cite{Naculich:2007nc,Naculich:1990pa,Naculich:1992uf},\cite{Naculich:1990hg,Naculich:1990bu,Fuchs:1989rv,Mlawer:1990uv,Naculich:2005tn,Nakanishi:1990hj} provided that $q=\exp[2\pi i /(N+K)]$ and $\theta=0$ mod $2\pi/(K+N)$. Another set of examples with similar themes is the left-right entanglement entropy (LREE) for WZW models on a circle\cite{PandoZayas:2014wsa,Das:2015oha,Schnitzer:2015gpa,Brehm:2015plf}, which is also related to the entropy of a (2+1) topological field theory (TQFT)\cite{Kitaev:2005dm,Dong:2008ft,Wen:2016snr}.

\section{R\'enyi and entanglement entropies}
The partition function of the 2d q-deformed YM theory on a genus g Riemann surface $\Sigma_g$ is given by (\ref{1.2}) \footnote{ An extensive exposition of the necessary representation theory for this paper is contained in ref. \cite{Naculich:2007nc}} Representations $\R$ of $\rm{\hat{U}(N)}_{K,N(K+N)}$ are defined by the representation 
\bea
\label{2.1}
(R,Q){\rm \quad of \quad  }  \rm{\widehat{SU}(N)_K}\times \rm{\hat{U}(N)}_{N(K+N)}
\eea
where
\bea
Q=r\,\,\,{\rm mod} \,\, \,N
\eea
with $r$ the number of boxes of the Young tableau associated to $\R$. The quadratic Casimir operator
\bea
C_2(\R)=C_2(R,Q)=NQ+T(\R)
\eea
and 
\bea
\label{2.5}
T(\R)=\sum_{i=1}^N
\bar{l}_i(\bar{l}_i-2i+1)
\eea
for an extended Young tableau $\R$ with row lengths $\bar{l}_i$, so that 
\bea
\label{2.4}
C_1(\R)=\sum_{i=1}^N\bar{l}_i
\eea

In order to compute the R\'enyi entropy by means of the replica trick, one replaces the base manifold $\Sigma_g$ by the n-sheeted ramified cover $\Sigma_n$ of $\Sigma_g$ with Euler characteristic (\ref{1.3}). The R\'enyi entropy for integer $n$ is 
\bea
\label{2.6}
S_n=\frac{1}{1-n}\ln \tr \rho_A^n
\eea
where the reduced density matrix $\rho_A=\tr_{\bar{A}}\rho$, with the trace taken over the complement $\bar{A}$. Then 
\bea
\label{2.7}
\tr\rho_A^n=\frac{Z_n}{Z^n}
\eea
and the EE is 
\bea
\label{2.8}
S_{EE}&=&-\lim_{n\to 1}\frac{\partial}{\partial n}\frac{Z_n}{Z^n} \\
&=&\ln Z-\frac 1Z \lim_{n\to 1}\frac{\partial}{\partial n}Z_n\nonumber 
\eea
Equations (\ref{1.2}), (\ref{1.3}) and (\ref{2.7}) give 
\bea
\label{2.9}
\tr \rho^n_A=\frac{\sum_{\R}[(\dim_q \mathcal{R})^{n\chi(\Sigma_g)-2l(n-1)}q^{-\frac{n}2pC_2(\mathcal{R})}e^{in\theta C_1(\mathcal{R})}]}{[\sum_{\R}(\dim_q \mathcal{R})^{\chi(\Sigma_g)}q^{-\frac{1}2pC_2(\mathcal{R})}e^{i\theta C_1(\mathcal{R})}]^n}
\eea
so that 
\bea
\label{2.10}
S_{EE}&=&\ln\,\left[\sum_{\R}(\dim_q \mathcal{R})^{\chi(\Sigma_g)}q^{-\frac{1}2pC_2(\mathcal{R})}e^{i\theta C_1(\mathcal{R})}\right]\nonumber \\
&&-\,\,\frac{\sum_{\R}[(\dim_q \mathcal{R})^{\chi(\Sigma_g)}q^{-\frac{1}2pC_2(\mathcal{R})}e^{i\theta C_1(\mathcal{R})}\ln\,[(\dim_q \mathcal{R})^{\chi(\Sigma_g)-2l}q^{-\frac{1}2pC_2(\mathcal{R})}e^{i\theta C_1(\mathcal{R})}]]}{\sum_{\R}(\dim_q \mathcal{R})^{\chi(\Sigma_g)}q^{-\frac{1}2pC_2(\mathcal{R})}e^{i\theta C_1(\mathcal{R})}}\nonumber \\
\eea
Equations (\ref{2.9}) and (\ref{2.10}) are the central results of this paper \footnote{In \cite{Gromov:2014kia} the EE is discussed for 2d non-Abelian pure gauge theory, an area preserving theory. Here we are considering topological gauge theories. We have greatly benefited from the considerations of \cite{Gromov:2014kia}.} Various specializations of the 2d qYM theory are obtained from (\ref{2.9}) and (\ref{2.10}) as we now detail.

When $q=\exp[2\pi i/(N+K)]$, $p\in \mathbb{Z}$\,, $\theta=\frac{\pi p(K+1)}{N+K}$ mod $\frac{2\pi}{N+K}$, $K$ odd, the sum (\ref{1.2}) is now restricted to Young tableaux with no more than $K$ columns\cite{Naculich:2007nc,Naculich:1990pa,Naculich:1992uf}. If further $\theta=2\pi t/(N+K)$, with $t \in \mathbb{Z}$, the quantities that appear in (\ref{1.2}) may be expressed in terms of the modular transformation matrices of  $\hat{U}(N)_{K,N(N+K)}$, namely\cite{Naculich:2007nc,Naculich:1990pa,Naculich:1992uf}
\bea
\label{2.11}
(\dim_q \mathcal{R})=(\dim_q R)=\left(\frac{S_{0R}}{S_{00}}\right)_{\widehat{SU}(N)_K}=\left(\frac{S_{0\R}}{S_{00}}\right)_{\hat{U}(N)_{K , N(N+K)}}
\eea
and
\bea
\label{2.12}
{\rm and} \quad q^{\frac12 C_2(\R)}=T_{\R \R}/T_{00}\,,
\eea
while $C_1(\R)$ is given by (\ref{2.4}). Then (\ref{2.11}) and (\ref{2.12}) are substituted into (\ref{2.9}) and (\ref{2.10}) to obtain the corresponding R\'enyi and entanglement entropies.

Finally, for $q=\exp[2\pi i/(N+K)]$, and $\theta=0$, the partition function may be expressed in terms of the level-K $U(N)$ CS theory on the Seifert manifold $\mathcal{M}_{(g,p)}$. [with Seifert framing]. Then (\ref{1.2}) reduces to\cite{Naculich:2007nc,Naculich:1990pa,Naculich:1992uf}
\bea
\label{2.13}
Z_{qYM}[U(N),e^{2\pi i/(N+K)},p, \theta=0]=T_{00}^p S_{00}^{2g-2}Z_{CS}[\mathcal{M}_{(g,p)},U(N),K] \! \! \quad (K \,\, \rm{ odd})\nonumber \\
\eea
where $\mathcal{M}{(g,p)}$ is the circle bundle over $\Sigma_g$ with first Chern class $p$. One then proceeds to the R\'enyi and EE by the substitution\cite{Gromov:2014kia}
\bea
\label{2.14}
2g-2=\chi(\Sigma_g)\to \chi(\Sigma_n)=n\chi(\Sigma_g)-2l(n-1) 
\eea
with $l\geq 1$, as in (\ref{1.3}). The appropriate restrictions of (\ref{2.7}) $-$ (\ref{2.10}) provide the R\'enyi and EE in terms of modular transformation matrices, and the Seifert manifold, as per (\ref{2.14}). The computations with these substitutions are straightforward, so that we omit the details.

\section{The Chern-Simons dual}

Since CS theory is a topological field theory, its observables are topological invariants of the three manifold $\mathcal{M}$ on which the theory is defined. The partition function for the theory depends only on $\mathcal{M}$, the gauge group $G$, the CS coupling $K$, and the framing of the manifold. Among other gauge invariant observables are Wilson lines defined by $\Tr_R P \exp(\oint A)$ around a closed path in $\mathcal{M}$, with the trace taken in some representation $R$ of the $G$. 

Consider CS theory on a Seifert manifold $\mathcal{M}(g,p)$, which is a circle bundle over $\Sigma_g$ with first Chern class $p$, together with the link consisting of $n-1$ circles carrying representations $R_2, \cdots, R_n$, all linked to a single circle carrying representation $R_1$. This has the expectation value in Seifert framing\cite{Naculich:2007nc,Naculich:1990pa,Naculich:1992uf}
\bea
\label{3.1}
W_{R_1\cdots R_n}[\mathcal{M}(g,p),G,K]=\sum_R (T_{RR})^{-p} S_{0R}^{2-n-2g}\prod_{i=1}^{n} S_{RR_i} 
\eea
where $T_{RR}$ and $S_{RR_i}$ are the modular transformations matrices of the $\hat{g}_K$ WZW model.

It has been shown\cite{Naculich:2007nc,Naculich:1990pa,Naculich:1992uf} that a certain class of Wilson line observables in $2d$ qYM theory with 
$q=\exp[2\pi i/(N+K)]$, are proportional to the links of CS theory on $\mathcal{M}(g,p)$. For $K$ odd these observables of $U(N)$ qYM
 theory may be expressed in terms of modular transformation matrices $\hat{U}(N)_{K , N(N+K)}$. That is\cite{Naculich:2007nc,Naculich:1990pa,Naculich:1992uf}
\bea
\label{3.2a}
W_{R_1\cdots R_n}[U(N),e^{2\pi i/(N+K)},p, 0]&\sim &\sum_R  S_{0R}^{2-2g-n}(T_{RR})^{-p} \prod_{i=1}^{n} S_{RR_i} \quad \!\! \textrm{(for $K$ odd)} \nonumber \\ \\
\label{3.2b}
&=&W_{R_1\cdots R_n}[\mathcal{M}(g,p),\, U(N),\,K]\nonumber 
\\
\eea
with Seifert framing.

In (\ref{3.2a}) the sum is restricted to $\R$ integrable, i.e. to a Young Tableaux with no more than $N $ rows and $K$ columns. Then the link described by (\ref{3.1}) for the CS theory provides the identity (\ref{3.2b}). The relation of $\R$ to $R$ is in (\ref{2.1}-\ref{2.4}). 

Equations (\ref{3.2a}-\ref{3.2b}), combined with (\ref{2.14}) allows us to compute the additional EE for the link observables described by (\ref{3.1})

We adopt the notation of \cite{Gromov:2014kia} to define the expectation value of an operator $X$ that is diagonal in $\R$, where 
\bea
\label{3.3}
X_\R=\langle\R |X|\R\rangle 
\eea
then
\bea
\langle X\rangle =\frac{\sum_{\R}[(\dim_q \mathcal{R})^{\chi(\Sigma_g)}q^{-\frac{1}2pC_2(\mathcal{R})}e^{i\theta C_1(\mathcal{R})}X_\R]}{\sum_{\R}(\dim_q \mathcal{R})^{\chi(\Sigma_g)}q^{-\frac{1}2pC_2(\mathcal{R})}e^{i\theta C_1(\mathcal{R})}}
\eea
Specialize this to (\ref{3.2a}-\ref{3.2b}), with $\theta=0$, and 
$\langle W_{\R_1\cdots \R_n}\rangle$ on the 2d qYM side and 
$\langle W_{R_1\cdots R_n}\rangle$ on the CS side. Define, together with (\ref{2.14}),(\ref{3.2a}) and (\ref{3.2b})
\bea
f(\R)=(\dim_q \mathcal{R})^{\chi(\Sigma_g)}q^{-\frac{1}2pC_2(\mathcal{R})}
\eea
then 
\bea
\label{3.6a}
\ln \langle W_{\R_1\cdots \R_n}\rangle &=&\ln [\sum_\R f(\R)W_{\R_1\cdots \R_n}]-\ln[\sum_\R f(\R)] \\
\label{3.6b}
&=&\ln [\sum_R f(R)W_{R_1\cdots R_n}]-\ln[\sum_R f(R)]
\eea
where (\ref{3.6a}) and (\ref{3.6b}) are for the 2d qYM side and CS side respectively.
Therefore the additional EE dual to the Wilson line observables described by (\ref{3.2a}-\ref{3.2b}) is given by (\ref{3.6a}-\ref{3.6b})

\section{Level-rank duality}

Restrict $N$ and $K$ both to odd values and $p$ even, so that both $\hat{U}(N)_{K , N(N+K)}$ and $\hat{U}(K)_{N , N(N+K)}$ are well defined. Then there is a level-rank dual map\cite{Naculich:2007nc,Naculich:1990pa,Naculich:1992uf} of the following: $\Tr \rho_A^n$, $S_{EE}$, $Z_{qYM}$, $W_{\R_1 \cdots \R_n}$ and $\Delta S_{EE}$.
For example,
\bea
\label{A.1}
W_{\R_1\cdots \R_n}[\mathcal{M}(g,p),\,U(N),\,K]=e^{i\pi p(KN+1)/12}W^{*}_{\tilde{\R}_1\cdots \tilde{\R}_n}[\mathcal{M}(g,p),\,U(N),\,K]
\eea
Analogous relations for (\ref{2.9}), (\ref{2.10}), (\ref{2.13}) and (\ref{3.6a}-\ref{3.6b}) are valid for the $\widehat{Sp}(n)_k$ WZW model, where $n=$ rank$Sp(n)$. That is, for $p$ even
\bea
\label{A.2}
W_{R_1\cdots R_n}[\mathcal{M}(g,p),\, Sp(n),\, k]=e^{i\pi pn k/6}W^{*}_{\tilde{R}_1\cdots \tilde{R}_n}[\mathcal{M}(g,p),\, Sp(n),\, k]
\eea
In equations (\ref{A.1}) and (\ref{A.2}) the representations $\tilde{R}$ is that obtained from $R$ with the rows and columns of the Young Tableux interchanged.

\section{Summary}

In this paper we examined a particular case where the topological R\'enyi and entanglement entropies can be computed in the presence of continuum gauge fields. In this context we considered the 2d q-deformed $U(N)$ Yang-Mills theory and its Chern-Simons dual, and particular Wilson loop link configurations.

It remains to be seen whether the example of this paper, and others, will be sufficient to provide insights for the treatment of R\'enyi and EE in theories with gauge fields without recourse to lattice\cite{Velytsky:2008rs,Buividovich:2008kq,Buividovich:2008gq,Nakagawa:2009jk,Klebanov:2011td,Eling:2013aqa,Agon:2013iva,Casini:2013rba,Donnelly:2011hn,Donnelly:2014gva,Casini:2014aia,Casini:2015dsg,Donnelly:2014fua,Donnelly:2015hxa,Huang:2014pfa,Ghosh:2015iwa,Aoki:2015bsa,Chen:2015kfa,Hung:2015fla,Radicevic:2014kqa,Radicevic:2015sza,Soni:2015yga,VanAcoleyen:2015ccp} or analogous regulators.  

\section*{Acknowledgments}

We are grateful to Cesar Ag\'on and Isaac Cohen for their invaluable assistance in preparing this paper. The research of H.J. Schnitzer is supported in part by the DOE by grant  DE-SC0009987

\newpage
\bibliographystyle{utphys}

\bibliography{emirefs}

\providecommand{\href}[2]{#2}\begingroup\raggedright\begin{thebibliography}{10}

\bibitem{Velytsky:2008rs}
A.~Velytsky, ``{Entanglement entropy in d+1 SU(N) gauge theory},''
  \href{http://dx.doi.org/10.1103/PhysRevD.77.085021}{{\em Phys. Rev.} {\bf
  D77} (2008)  085021},
\href{http://arxiv.org/abs/0801.4111}{{\tt arXiv:0801.4111 [hep-th]}}.

\bibitem{Buividovich:2008kq}
P.~V. Buividovich and M.~I. Polikarpov, ``{Numerical study of entanglement
  entropy in SU(2) lattice gauge theory},''
  \href{http://dx.doi.org/10.1016/j.nuclphysb.2008.04.024}{{\em Nucl. Phys.}
  {\bf B802} (2008)  458--474},
\href{http://arxiv.org/abs/0802.4247}{{\tt arXiv:0802.4247 [hep-lat]}}.

\bibitem{Buividovich:2008gq}
P.~V. Buividovich and M.~I. Polikarpov, ``{Entanglement entropy in gauge
  theories and the holographic principle for electric strings},''
  \href{http://dx.doi.org/10.1016/j.physletb.2008.10.032}{{\em Phys. Lett.}
  {\bf B670} (2008)  141--145},
\href{http://arxiv.org/abs/0806.3376}{{\tt arXiv:0806.3376 [hep-th]}}.

\bibitem{Nakagawa:2009jk}
Y.~Nakagawa, A.~Nakamura, S.~Motoki, and V.~I. Zakharov, ``{Entanglement
  entropy of SU(3) Yang-Mills theory},'' {\em PoS} {\bf LAT2009} (2009)  188,
\href{http://arxiv.org/abs/0911.2596}{{\tt arXiv:0911.2596 [hep-lat]}}.

\bibitem{Klebanov:2011td}
I.~R. Klebanov, S.~S. Pufu, S.~Sachdev, and B.~R. Safdi, ``{Entanglement
  Entropy of 3-d Conformal Gauge Theories with Many Flavors},''
  \href{http://dx.doi.org/10.1007/JHEP05(2012)036}{{\em JHEP} {\bf 05} (2012)
  036},
\href{http://arxiv.org/abs/1112.5342}{{\tt arXiv:1112.5342 [hep-th]}}.

\bibitem{Eling:2013aqa}
C.~Eling, Y.~Oz, and S.~Theisen, ``{Entanglement and Thermal Entropy of Gauge
  Fields},'' \href{http://dx.doi.org/10.1007/JHEP11(2013)019}{{\em JHEP} {\bf
  11} (2013)  019},
\href{http://arxiv.org/abs/1308.4964}{{\tt arXiv:1308.4964 [hep-th]}}.

\bibitem{Agon:2013iva}
C.~A. Agon, M.~Headrick, D.~L. Jafferis, and S.~Kasko, ``{Disk entanglement
  entropy for a Maxwell field},''
  \href{http://dx.doi.org/10.1103/PhysRevD.89.025018}{{\em Phys. Rev.} {\bf
  D89} (2014) no.~2, 025018},
\href{http://arxiv.org/abs/1310.4886}{{\tt arXiv:1310.4886 [hep-th]}}.

\bibitem{Casini:2013rba}
H.~Casini, M.~Huerta, and J.~A. Rosabal, ``{Remarks on entanglement entropy for
  gauge fields},'' \href{http://dx.doi.org/10.1103/PhysRevD.89.085012}{{\em
  Phys. Rev.} {\bf D89} (2014) no.~8, 085012},
\href{http://arxiv.org/abs/1312.1183}{{\tt arXiv:1312.1183 [hep-th]}}.

\bibitem{Donnelly:2011hn}
W.~Donnelly, ``{Decomposition of entanglement entropy in lattice gauge
  theory},'' \href{http://dx.doi.org/10.1103/PhysRevD.85.085004}{{\em Phys.
  Rev.} {\bf D85} (2012)  085004},
\href{http://arxiv.org/abs/1109.0036}{{\tt arXiv:1109.0036 [hep-th]}}.

\bibitem{Donnelly:2014gva}
W.~Donnelly, ``{Entanglement entropy and nonabelian gauge symmetry},''
  \href{http://dx.doi.org/10.1088/0264-9381/31/21/214003}{{\em Class. Quant.
  Grav.} {\bf 31} (2014) no.~21, 214003},
\href{http://arxiv.org/abs/1406.7304}{{\tt arXiv:1406.7304 [hep-th]}}.

\bibitem{Casini:2014aia}
H.~Casini and M.~Huerta, ``{Entanglement entropy for a Maxwell field: Numerical
  calculation on a two dimensional lattice},''
  \href{http://dx.doi.org/10.1103/PhysRevD.90.105013}{{\em Phys. Rev.} {\bf
  D90} (2014) no.~10, 105013},
\href{http://arxiv.org/abs/1406.2991}{{\tt arXiv:1406.2991 [hep-th]}}.

\bibitem{Casini:2015dsg}
H.~Casini and M.~Huerta, ``{Entanglement entropy of a Maxwell field on the
  sphere},''
\href{http://arxiv.org/abs/1512.06182}{{\tt arXiv:1512.06182 [hep-th]}}.

\bibitem{Donnelly:2014fua}
W.~Donnelly and A.~C. Wall, ``{Entanglement entropy of electromagnetic edge
  modes},'' \href{http://dx.doi.org/10.1103/PhysRevLett.114.111603}{{\em Phys.
  Rev. Lett.} {\bf 114} (2015) no.~11, 111603},
\href{http://arxiv.org/abs/1412.1895}{{\tt arXiv:1412.1895 [hep-th]}}.

\bibitem{Donnelly:2015hxa}
W.~Donnelly and A.~C. Wall, ``{Geometric entropy and edge modes of the
  electromagnetic field},''
\href{http://arxiv.org/abs/1506.05792}{{\tt arXiv:1506.05792 [hep-th]}}.

\bibitem{Huang:2014pfa}
K.-W. Huang, ``{Central Charge and Entangled Gauge Fields},''
  \href{http://dx.doi.org/10.1103/PhysRevD.92.025010}{{\em Phys. Rev.} {\bf
  D92} (2015) no.~2, 025010},
\href{http://arxiv.org/abs/1412.2730}{{\tt arXiv:1412.2730 [hep-th]}}.

\bibitem{Ghosh:2015iwa}
S.~Ghosh, R.~M. Soni, and S.~P. Trivedi, ``{On The Entanglement Entropy For
  Gauge Theories},'' \href{http://dx.doi.org/10.1007/JHEP09(2015)069}{{\em
  JHEP} {\bf 09} (2015)  069},
\href{http://arxiv.org/abs/1501.02593}{{\tt arXiv:1501.02593 [hep-th]}}.

\bibitem{Aoki:2015bsa}
S.~Aoki, T.~Iritani, M.~Nozaki, T.~Numasawa, N.~Shiba, and H.~Tasaki, ``{On the
  definition of entanglement entropy in lattice gauge theories},''
  \href{http://dx.doi.org/10.1007/JHEP06(2015)187}{{\em JHEP} {\bf 06} (2015)
  187},
\href{http://arxiv.org/abs/1502.04267}{{\tt arXiv:1502.04267 [hep-th]}}.

\bibitem{Chen:2015kfa}
J.-W. Chen, S.-H. Dai, and J.-Y. Pang, ``{Strong Coupling Expansion of the
  Entanglement Entropy of Yang-Mills Gauge Theories},''
\href{http://arxiv.org/abs/1503.01766}{{\tt arXiv:1503.01766 [hep-th]}}.

\bibitem{Hung:2015fla}
L.-Y. Hung and Y.~Wan, ``{Revisiting Entanglement Entropy of Lattice Gauge
  Theories},'' \href{http://dx.doi.org/10.1007/JHEP04(2015)122}{{\em JHEP} {\bf
  04} (2015)  122},
\href{http://arxiv.org/abs/1501.04389}{{\tt arXiv:1501.04389 [hep-th]}}.

\bibitem{Radicevic:2014kqa}
D.~Radicevic, ``{Notes on Entanglement in Abelian Gauge Theories},''
\href{http://arxiv.org/abs/1404.1391}{{\tt arXiv:1404.1391 [hep-th]}}.

\bibitem{Radicevic:2015sza}
D.~Radicevic, ``{Entanglement in Weakly Coupled Lattice Gauge Theories},''
  \href{http://dx.doi.org/10.1007/JHEP04(2016)163}{{\em JHEP} {\bf 04} (2016)
  163},
\href{http://arxiv.org/abs/1509.08478}{{\tt arXiv:1509.08478 [hep-th]}}.

\bibitem{Soni:2015yga}
R.~M. Soni and S.~P. Trivedi, ``{Aspects of Entanglement Entropy for Gauge
  Theories},'' \href{http://dx.doi.org/10.1007/JHEP01(2016)136}{{\em JHEP} {\bf
  01} (2016)  136},
\href{http://arxiv.org/abs/1510.07455}{{\tt arXiv:1510.07455 [hep-th]}}.

\bibitem{VanAcoleyen:2015ccp}
K.~Van~Acoleyen, N.~Bultinck, J.~Haegeman, M.~Marien, V.~B. Scholz, and
  F.~Verstraete, ``{The entanglement of distillation for gauge theories},''
\href{http://arxiv.org/abs/1511.04369}{{\tt arXiv:1511.04369 [quant-ph]}}.

\bibitem{1990NuPhB.332..583N}
S.~G. {Naculich} and H.~J. {Schnitzer},
  \href{http://dx.doi.org/10.1016/0550-3213(90)90003-V}{``{Constructive methods
  for higher-genus correlation functions of level-one simply-laced WZW
  models},''{\em Nuclear Physics B} {\bf 332} (Mar., 1990)  583--628}.

\bibitem{Schnitzer:2015ira}
H.~J. Schnitzer, ``{R\'enyi Entropy for the $\sun1$ WZW model on the torus},''
\href{http://arxiv.org/abs/1510.05993}{{\tt arXiv:1510.05993 [hep-th]}}.

\bibitem{Schnitzer:2015iip}
H.~J. Schnitzer, ``{in preparation},''.

\bibitem{PandoZayas:2014wsa}
L.~A. Pando~Zayas and N.~Quiroz, ``{Left-Right Entanglement Entropy of Boundary
  States},'' \href{http://dx.doi.org/10.1007/JHEP01(2015)110}{{\em JHEP} {\bf
  01} (2015)  110},
\href{http://arxiv.org/abs/1407.7057}{{\tt arXiv:1407.7057 [hep-th]}}.

\bibitem{Das:2015oha}
D.~Das and S.~Datta, ``{Universal features of left-right entanglement
  entropy},'' \href{http://dx.doi.org/10.1103/PhysRevLett.115.131602}{{\em
  Phys. Rev. Lett.} {\bf 115} (2015) no.~13, 131602},
\href{http://arxiv.org/abs/1504.02475}{{\tt arXiv:1504.02475 [hep-th]}}.

\bibitem{Schnitzer:2015gpa}
H.~J. Schnitzer, ``{Left-Right Entanglement Entropy, D-Branes, and Level-rank
  duality},''
\href{http://arxiv.org/abs/1505.07070}{{\tt arXiv:1505.07070 [hep-th]}}.

\bibitem{Brehm:2015plf}
E.~M. Brehm, I.~Brunner, D.~Jaud, and C.~Schmidt-Colinet, ``{Entanglement and
  topological interfaces},''
\href{http://arxiv.org/abs/1512.05945}{{\tt arXiv:1512.05945 [hep-th]}}.

\bibitem{Gromov:2014kia}
A.~Gromov and R.~A. Santos, ``{Entanglement Entropy in 2D Non-abelian Pure
  Gauge Theory},'' \href{http://dx.doi.org/10.1016/j.physletb.2014.08.023}{{\em
  Phys. Lett.} {\bf B737} (2014)  60--64},
\href{http://arxiv.org/abs/1403.5035}{{\tt arXiv:1403.5035 [hep-th]}}.

\bibitem{Naculich:2007nc}
S.~G. Naculich and H.~J. Schnitzer, ``{Level-rank duality of the U(N) WZW
  model, Chern-Simons theory, and 2-D qYM theory},''
  \href{http://dx.doi.org/10.1088/1126-6708/2007/06/023}{{\em JHEP} {\bf 06}
  (2007)  023},
\href{http://arxiv.org/abs/hep-th/0703089}{{\tt arXiv:hep-th/0703089
  [HEP-TH]}}.

\bibitem{Naculich:1990pa}
S.~G. Naculich, H.~A. Riggs, and H.~J. Schnitzer, ``{Group Level Duality in
  {WZW} Models and {Chern-Simons} Theory},''
\href{http://dx.doi.org/10.1016/0370-2693(90)90623-E}{{\em Phys. Lett.} {\bf
  B246} (1990)  417--422}.

\bibitem{Naculich:1992uf}
S.~G. Naculich, H.~A. Riggs, and H.~J. Schnitzer, ``{Simple current symmetries,
  rank level duality, and linear skein relations for Chern-Simons graphs},''
  \href{http://dx.doi.org/10.1016/0550-3213(93)90022-H}{{\em Nucl. Phys.} {\bf
  B394} (1993)  445--508},
\href{http://arxiv.org/abs/hep-th/9205082}{{\tt arXiv:hep-th/9205082
  [hep-th]}}.

\bibitem{Aganagic:2004js}
M.~Aganagic, H.~Ooguri, N.~Saulina, and C.~Vafa, ``{Black holes, q-deformed 2d
  Yang-Mills, and non-perturbative topological strings},''
  \href{http://dx.doi.org/10.1016/j.nuclphysb.2005.02.035}{{\em Nucl. Phys.}
  {\bf B715} (2005)  304--348},
\href{http://arxiv.org/abs/hep-th/0411280}{{\tt arXiv:hep-th/0411280
  [hep-th]}}.

\bibitem{deHaro:2005rn}
S.~de~Haro, ``{A Note on knot invariants and q-deformed 2-D Yang-Mills},''
  \href{http://dx.doi.org/10.1016/j.physletb.2006.01.014}{{\em Phys. Lett.}
  {\bf B634} (2006)  78--83},
\href{http://arxiv.org/abs/hep-th/0509167}{{\tt arXiv:hep-th/0509167
  [hep-th]}}.

\bibitem{Blau:2006gh}
M.~Blau and G.~Thompson, ``{Chern-Simons theory on S1-bundles: Abelianisation
  and q-deformed Yang-Mills theory},''
  \href{http://dx.doi.org/10.1088/1126-6708/2006/05/003}{{\em JHEP} {\bf 05}
  (2006)  003},
\href{http://arxiv.org/abs/hep-th/0601068}{{\tt arXiv:hep-th/0601068
  [hep-th]}}.

\bibitem{Marino:2001re}
M.~Marino and C.~Vafa, ``{Framed knots at large N},'' {\em Contemp. Math.} {\bf
  310}  .

\bibitem{Marino:2004uf}
M.~Marino, ``{Chern-Simons theory and topological strings},''
  \href{http://dx.doi.org/10.1103/RevModPhys.77.675}{{\em Rev. Mod. Phys.} {\bf
  77} (2005)  675--720},
\href{http://arxiv.org/abs/hep-th/0406005}{{\tt arXiv:hep-th/0406005
  [hep-th]}}.

\bibitem{Blau:1993hj}
M.~Blau and G.~Thompson, ``{Lectures on 2-d gauge theories: Topological aspects
  and path integral techniques},'' in {\em {Summer School in High-energy
  Physics and Cosmology (Includes Workshop on Strings, Gravity, and Related
  Topics 29-30 Jul 1993) Trieste, Italy, June 14-July 30, 1993}}.
\newblock 1993.
\newblock
\href{http://arxiv.org/abs/hep-th/9310144}{{\tt arXiv:hep-th/9310144
  [hep-th]}}.
\newblock

\bibitem{Kitaev:2005dm}
A.~Kitaev and J.~Preskill, ``{Topological entanglement entropy},''
  \href{http://dx.doi.org/10.1103/PhysRevLett.96.110404}{{\em Phys. Rev. Lett.}
  {\bf 96} (2006)  110404},
\href{http://arxiv.org/abs/hep-th/0510092}{{\tt arXiv:hep-th/0510092
  [hep-th]}}.

\bibitem{Dong:2008ft}
S.~Dong, E.~Fradkin, R.~G. Leigh, and S.~Nowling, ``{Topological Entanglement
  Entropy in Chern-Simons Theories and Quantum Hall Fluids},''
  \href{http://dx.doi.org/10.1088/1126-6708/2008/05/016}{{\em JHEP} {\bf 05}
  (2008)  016},
\href{http://arxiv.org/abs/0802.3231}{{\tt arXiv:0802.3231 [hep-th]}}.

\bibitem{Wen:2016snr}
X.~Wen, S.~Matsuura, and S.~Ryu, ``{Edge theory approach to topological
  entanglement entropy, mutual information and entanglement negativity in
  Chern-Simons theories},''
\href{http://arxiv.org/abs/1603.08534}{{\tt arXiv:1603.08534
  [cond-mat.mes-hall]}}.

\bibitem{Naculich:1990hg}
S.~G. Naculich and H.~J. Schnitzer, ``{Duality Between SU($N$)-k and SU(k)-$N$
  {WZW} Models},''
\href{http://dx.doi.org/10.1016/0550-3213(90)90380-V}{{\em Nucl. Phys.} {\bf
  B347} (1990)  687--742}.

\bibitem{Naculich:1990bu}
S.~G. Naculich and H.~J. Schnitzer, ``{Duality Relations Between SU($N$)-k and
  SU(k)-$N$ {WZW} Models and Their Braid Matrices},''
\href{http://dx.doi.org/10.1016/0370-2693(90)90061-A}{{\em Phys. Lett.} {\bf
  B244} (1990)  235--240}.

\bibitem{Fuchs:1989rv}
J.~Fuchs and P.~van Driel, ``{Some Symmetries of Quantum Dimensions},''
\href{http://dx.doi.org/10.1063/1.528673}{{\em J. Math. Phys.} {\bf 31} (1990)
  1770--1775}.

\bibitem{Mlawer:1990uv}
E.~J. Mlawer, S.~G. Naculich, H.~A. Riggs, and H.~J. Schnitzer, ``{Group level
  duality of WZW fusion coefficients and Chern-Simons link observables},''
\href{http://dx.doi.org/10.1016/0550-3213(91)90110-J}{{\em Nucl. Phys.} {\bf
  B352} (1991)  863--896}.

\bibitem{Naculich:2005tn}
S.~G. Naculich and H.~J. Schnitzer, ``{Level-rank duality of D-branes on the
  SU(N) group manifold},''
  \href{http://dx.doi.org/10.1016/j.nuclphysb.2006.01.041}{{\em Nucl. Phys.}
  {\bf B740} (2006)  181--194},
\href{http://arxiv.org/abs/hep-th/0511083}{{\tt arXiv:hep-th/0511083
  [hep-th]}}.

\bibitem{Nakanishi:1990hj}
T.~Nakanishi and A.~Tsuchiya, ``{Level rank duality of WZW models in conformal
  field theory},''
\href{http://dx.doi.org/10.1007/BF02101097}{{\em Commun. Math. Phys.} {\bf 144}
  (1992)  351--372}.

\end{thebibliography}\endgroup

\end{document}